\documentclass{article}
\usepackage{latexsym}
\usepackage{xspace}
\usepackage[latin1]{inputenc}
\usepackage{natbib}
\setcitestyle{authoryear,open={(},close={)}}
\usepackage{epsfig}
\usepackage{amsmath}

\newtheorem{dfn}{Definition}

\newtheorem{ex}{Example}[section]

\newtheorem{property}{Property}[section]

\newtheorem{alg}{Algorithm}

\newcommand{\bmu}{\boldsymbol{\mu}}

\newcommand{\bone}{\boldsymbol{1}}

\newcommand{\ba}{\boldsymbol{a}}

\newcommand{\bx}{\boldsymbol{x}}

\newcommand{\by}{\boldsymbol{y}}
\newcommand{\bz}{\boldsymbol{z}}

\newcommand{\bg}{\boldsymbol{g}}

\newcommand{\bSigma}{\boldsymbol{\Sigma}}
\newcommand{\be}{\boldsymbol{e}}
\newcommand{\bzero}{\boldsymbol{0}}

\title{Likelihood Estimation with Incomplete Array Variate Observations}

\author{Deniz Akdemir\\ Plant Breeding and Genetics\\ Cornell University, Ithaca, NY}

\begin{document}
\maketitle

\begin{abstract}
Missing data present an important challenge when dealing with high dimensional data arranged in the form of an array. In this paper, we propose methods for estimation of the parameters of array variate normal probability model from partially observed multi-way data.  The methods developed here are useful for missing data imputation,  estimation of mean and covariance parameters for multi-way data. A multi-way semi-parametric mixed effects model that allows separation of multi-way mean and covariance effects is also defined, and an efficient algorithm for estimation based on the spectral decompositions of the covariance parameters is recommended. We demonstrate our methods with simulations and real life data involving the estimation of genotype and environment interaction effects on possibly correlated traits. 
\end{abstract}


\maketitle

\section{Introduction}

A vector is a one way array, a matrix is a two way array, by stacking matrices we obtain three way arrays, etc, ... Array variate random variables up to two dimensions has been studied intensively in \cite{gupta2000matrix} and by many others. For arrays observations of $3,$ $4$ or in general $i$ dimensions probability models with Kronecker delta covariance structure have been proposed very recently in (\cite{AKDEMIRjalgstat}, \cite{srivastava2008estimation} and \cite{ohlson2011multilinear}).  The estimation and inference for the parameters of the array normal distribution with Kronecker delta covariance structure, based on a random sample of fully observed arrays $\{\widetilde{X}_1, \widetilde{X}_2, \ldots, \widetilde{X}_N\},$ can been accomplished by maximum likelihood estimation (\cite{srivastava2008models}, \cite{AKDEMIRjalgstat}, \cite{srivastava2008estimation} and \cite{ohlson2011multilinear}) or by Bayesian estimation (\cite{hoff2011hierarchical}). 

Array variate random variables are mainly useful for multiply labeled random variables that can naturally be arranged in array form. Some examples include response from multi-factor experiments, two-three dimensional image-video data, spatial-temporal data, repeated measures data.  It is true that any array data can also be represented uniquely in vector form, and a general covariance structure can be assumed for this vector representation. However, the models with the Kronecker structure far more parsimonious. 

The array variate data models and the estimation techniques we have mentioned above assume that we have a random sample of fully observed arrays. However, in practice most array data come with many missing cells. The main purpose of this article is to develop likelihood-based methods for estimation and inference for a class of array random variables when we only have partially observed arrays in the random sample. 

Another novelty in this article involves the definition and development of a multiway mixed effects model. This model is useful for analyzing multiway response variables that depend on separable effects and through it we can incorporate the known covariance structures along some dimensions of the response, and we can estimate the unknown mean and covariance components. 

The array variate mixed models can be  used  to fit Gaussian process regression models with multiway data. Using the explanatory information that describe levels related to the dimension of an array, we can calculate a kernel matrix for that dimension. The shrinkage parameters related to a kernel along a dimension can be estimated using likelihood based methods. Similarly, the covariance for the dimensions with no explanatory information can also be estimated.  We illustrate this with two examples where we calculate and use kernel matrices based on genetic information in the form of genomewide markers. 

The remaining of the article is organized as follows: In Section \ref{sec2}, we  introduce the normal model for array variables. In Section \ref{sec3}, we introduce the updating equations for parameter estimation and missing data imputation. In Section \ref{sec4}, the basic algorithm is introduced. Section \ref{sec5}, we define a semi-parametric  array variate mixed model with Kronecker covariance structure, and an efficient algorithm for the estimation of variance components is described. In section \ref{sec6}, we study the principal component analysis for the array care. Examples illustrating the use of these methods are provided in Section \ref{sec7}, followed by our conclusions in Section \ref{sec8}.

\section{Array Normal Random Variable}\label{sec2}

The family of normal densities with Kronecker delta covariance structure are given by \small \begin{equation}\label{arraymodel}\phi(\widetilde{X}; \widetilde{\mathcal{M}},\mathcal{A}_1,\mathcal{A}_2,\ldots \mathcal{A}_i)=\frac{\exp{(-\frac{1}{2}\|{(\mathcal{A}_1^{-1})^1 (\mathcal{A}_2^{-1})^2 \ldots (\mathcal{A}_i^{-1})^i(\widetilde{X}-\widetilde{\mathcal{M}})}\|^2)}}{(2\pi)^{(\prod_j m_j)/2}|\mathcal{A}_1|^{\prod_{j\neq 1}{m_j}} |\mathcal{A}_2|^{\prod_{j\neq 2}{m_j}} \ldots |\mathcal{A}_i|^{\prod_{j\neq i}{m_j}}}\end{equation} \normalsize where $\mathcal{A}_1, \mathcal{A}_2,\ldots,\mathcal{A}_i$ are non-singular matrices of orders $m_1, m_2,\ldots, m_i;$ the R-Matrix multiplication (\cite{rauhala2002array}) which generalizes the matrix multiplication (array multiplication in two dimensions) to the case of $k$-dimensional arrays is defined element wise as
$$((\mathcal{A}_1)^1 (\mathcal{A}_2)^2 \ldots (\mathcal{A}_i)^i\widetilde{X}_{m_1 \times m_2 \times \ldots \times m_i})_{q_1q_2\ldots q_i}$$ $$=\sum_{r_1=1}^{m_1}(\mathcal{A}_1)_{q_1r_1}\sum_{r_2=1}^{m_2}(\mathcal{A}_2)_{q_2r_2}\sum_{r_3=1}^{m_3}(\mathcal{A}_3)_{q_3r_3}\ldots \sum_{r_i=1}^{m_i}(\mathcal{A}_i)_{q_ir_i}(\widetilde{X})_{r_1r_2\ldots r_i}$$ and the square norm of $\widetilde{X}_{m_1 \times m_2 \times\ldots m_i}$ is defined as $$\|\widetilde{X}\|^2=\sum_{j_1=1}^{m_1}\sum_{j_2=1}^{m_2}\ldots\sum_{j_i=1}^{m_i}((\widetilde{X})_{j_1j_2\ldots j_i})^2.$$  Note that R-Matrix multiplication is sometimes referred to as the Tucker product or $n-$mode product (\cite{kolda2006multilinear}). 

An important operation with an array is the matricization (also known as unfolding or flattening) operation, it is the process of arranging the elements of an array in a matrix. Matricization of an array of dimensions $m_1, \times m_2, \ldots, m_i$  along its $k$th dimension is obtained by stacking the $m_k$ dimensional column vectors along the $k$th in the order of the levels of the other dimensions and results in a $m_k\times \prod_{j\neq k} m_j$ matrix. 

The operator $rvec$ describes the relationship between $\widetilde{X}_{m_1 \times m_2 \times \ldots m_i}$ and its mono-linear form $\bx_{m_1m_2\ldots m_i\times 1}.$ $rvec( \widetilde{X}_{m_1 \times m_2 \times \ldots m_i})=\bx_{m_1m_2\ldots m_i\times 1}$ where $\bx$ is the column vector obtained by stacking the elements of the array $\widetilde{X}$ in the order of its dimensions; i.e., $(\widetilde{X})_{j_1 j_2 \ldots j_i}=(\bx)_j$ where $j=(j_i-1)m_{i-1}m_{i-2}\ldots m_1+(j_i-2)m_{i-2}m_{i-3}\ldots m_1+\ldots+(j_2-1)m_1+j_1.$

The following are very useful properties of the array normal variable with Kronecker Delta covariance structure (\cite{AKDEMIRjalgstat}).
\begin{property}\label{property1}If $\widetilde{X}$ $\sim$ $\phi(\widetilde{X};$  $\widetilde{\mathcal{M}},$ $\mathcal{A}_1,$ $\mathcal{A}_2,$ $\ldots$ $ \mathcal{A}_i)$ then $rvec(\widetilde{X})$ $\sim$ $ \phi(rvec(\widetilde{X});$ $ rvec(\widetilde{\mathcal{M}}),$ $\mathcal{A}_i$ $\otimes$ $\ldots$ $\otimes$ $ \mathcal{A}_2$ $\otimes \mathcal{A}_1).$ \end{property}

\begin{property}If $\widetilde{X}\sim\phi(\widetilde{X}; \widetilde{\mathcal{M}},\mathcal{A}_1,\mathcal{A}_2,\ldots \mathcal{A}_i)$ then $E(rvec(\widetilde{X}))=rvec(\widetilde{\mathcal{M}})$ and $cov(rvec(\widetilde{X}))=(\mathcal{A}_i\otimes\ldots\otimes \mathcal{A}_2\otimes \mathcal{A}_1)(\mathcal{A}_i\otimes\ldots\otimes \mathcal{A}_2\otimes \mathcal{A}_1)'.$ \end{property}

In the remaining of this paper we will assume that the matrices $\mathcal{A}_i$ are unique square roots (for example, eigenvalue or Chelosky decompositions) of the positive definite matrices $\bSigma_i$ for $i=1,2,\ldots,i$ and we will put $\Lambda=\bSigma_i\otimes\ldots\otimes \bSigma_2\otimes \bSigma_1$  $=(\mathcal{A}_i\otimes\ldots\otimes \mathcal{A}_2\otimes \mathcal{A}_1)(\mathcal{A}_i\otimes\ldots\otimes \mathcal{A}_2\otimes \mathcal{A}_1)'$ for the overall covariance matrix.

We also use the following notation: 
\begin{itemize}
\item \[(A)^k\widetilde{X}\equiv (I)^1(I)^2\ldots (I)^{k-1}(A)^k(I)^{k+1}\ldots (I)^i\widetilde{X}.\]
\item For vectors $\ba_k,$ $k=1,2,\ldots, i,$ \[(\ba_1)^1(\ba_2)^2 \ldots (\ba_i)^i 1\equiv (\ba_1)^1(\ba_2)^2 \ldots (\ba_i)^i \widetilde{1}_{1\times 1 \times \ldots \times 1}.\]
\item Matricization of $\widetilde{X}$ along $k$th dimension: $X_{(k)}$
\item For ease of notation, when the dimensions are evident from the context, we have used $\bzero$ to stand for the zero matrix with appropriate dimensions.
\item A vector of ones: $\bone.$
\end{itemize}

\section{Updates for missing values and the parameters}\label{sec3}

Using linear predictors for the purpose of imputing missing values in multivariate normal data dates back at least as far as (\cite{anderson1957maximum}). The EM algorithm (\cite{dempster1977maximum}) is usually utilized for multivariate normal distribution with missing data. The EM method goes back to (\cite{orchard1972missing}) and (\cite{beale1975missing}). \cite{trawinski1964maximum} and \cite{hartley1971analysis} developed the Fisher scoring algorithm for incomplete multivariate normal data. The notation and the algorithms described in this section were adopted from \cite{jorgensen2012efficient}.  

Let $\bx$ be a $k$ dimensional observation vector which is partitioned as \[\left[ \begin{array}{c}
R \\
M  \end{array} \right]\bx= \left[ \begin{array}{c}
\bx_r  \\
\bx_m  \end{array} \right]\] where $\bx_r$ and $\bx_m$ represent the vector of observed values and the missing observations correspondingly. Here \[\left[ \begin{array}{c}
R \\
M  \end{array} \right]\] is an orthogonal permutation matrix of zeros and ones and \[\bx=\left[ \begin{array}{c}
R \\
M  \end{array} \right]'\left[ \begin{array}{c}
\bx_r  \\
\bx_m  \end{array} \right].\]

The the mean vector and the covariance matrix of $\left[\begin{array}{c}
\bx_r  \\
\bx_m  \end{array} \right]$ are given by 

\[ \left[ \begin{array}{c}
R \\
M  \end{array} \right]E(\bx) = \left[ \begin{array}{c}
\bmu_r  \\
\bmu_m  \end{array} \right]
\]

and

\[ \left[ \begin{array}{c}
R \\
M  \end{array} \right]cov(\bx)\left[ \begin{array}{c}
R \\
M  \end{array} \right]' = \left[ \begin{array}{cc}
\bSigma_{rr} & \bSigma_{rm} \\
\bSigma_{mr} & \bSigma_{mm} \end{array} \right]
\] correspondingly.

Let $\widetilde{X}_1,\widetilde{X}_2,\ldots,\widetilde{X}_N$ be a random sample of array observations from the distribution with density $\phi(\widetilde{X}; \widetilde{\mathcal{M}},\mathcal{A}_1,\mathcal{A}_2,\ldots \mathcal{A}_i).$ Let the current  values of the parameters be $\widetilde{\mathcal{M}}^t, \mathcal{A}^t_1,\mathcal{A}^t_2,\ldots \mathcal{A}^t_i.$ 

The mean of the conditional distribution of $rvec({\widetilde{X}_l})$ given the estimates of parameters at time $t$ can be obtained using \begin{equation}\label{eq:blupforx}rvec(\widehat{\widetilde{X}_l}^t)=rvec{\widetilde{\mathcal{M}}^t}+\Lambda^t R'_l(R_l\Lambda^t R'_l)^{-1}(R_l\bx_{l}-R_l rvec(\widetilde{\mathcal{M}}^t))\end{equation}
 where $\bx_{l}=rvec({\widetilde{X}_l})$ and $R_l$ is the permutation matrix such that $\bx_{rl}=R_l \bx_{l}.$
The updating equation of the parameter $\widetilde{\mathcal{M}}$ is given by
\begin{eqnarray}\label{em1}
 rvec(\widetilde{\mathcal{M}}^{t+1}) &=& \frac{1}{N}\sum_{l=1}^{N}rvec(\widehat{\widetilde{X}_l}^t).
\end{eqnarray}
 
To update the covariance matrix along the kth dimension calculate \[\widetilde{Z}=(\mathcal{A}_1^{-1})^1 (\mathcal{A}_2^{-1})^2 \ldots (\mathcal{A}_{k-1}^{-1})^{k-1}(I_{m_k})^k(\mathcal{A}_{k+1}^{-1})^{k+1}\ldots(\mathcal{A}_i^{-1})^i(\widehat{\widetilde{X}}^t-\widetilde{\mathcal{M}})\] using the most recent estimates of the parameters. Assuming that the values of the parameter values are correct we can write,  $\widetilde{Z}\sim$ $\phi(\widetilde{Z};$  $\widetilde{0},$ $I_{m_1},$ $I_{m_2},$ $\ldots,$ $I_{m_{k-1}},$ $\mathcal{A}_k,$ $I_{m_{k+1}}$,$\ldots,$ $I_{m_i}),$ i.e.,  $Z_{(k)}$ $\sim$ $\phi(Z_{(k)};$ $\bzero_{m_k\times\prod_{j\neq k}m_j},$ $\mathcal{A}_k,$ $I_{\prod_{j\neq k}m_j})$ where$Z_{(k)}$ denotes the $m_k \times \prod_{j\neq k} m_j$ matrix obtained by stacking the elements of $\widetilde{Z}$ along the $k$th dimension. Therefore, $(Z_{(k)1},$ $Z_{(k)2},$ $\ldots,$ $Z_{(k)N})$ $=(\bz_1,$ $\bz_2,$ $...$ $\bz_{N\prod_{j\neq k}m_j})$ can be treated as a random sample of size $N\prod_{j\neq k}m_j$ from the $m_k$-variate normal distribution with mean zero and covariance $\bSigma_k=\mathcal{A}_kA'_k.$ An update for $\bSigma_k$ can be obtained by calculating the sample covariance matrix for $Z_{(k)}:$
\begin{eqnarray}\label{em12}
\widehat{\bSigma_k}^{t+1} &=& \frac{1}{N\prod_{j\neq k}m_j}\sum_{q=1}^{N\prod_{j\neq k}m_j}Z_{(k)q}Z'_{(k)q}.
\end{eqnarray}
\section{Flip-Flop Algorithm for Incomplete Arrays}\label{sec4}

Inference about the parameters of the model in (\ref{arraymodel})  for the matrix variate case has been considered in the statistical literature (\cite{roy2003tests}, \cite{roy2008likelihood}, \cite{lu2005likelihood}, \cite{srivastava2008models}, etc.). The Flip-Flop Algorithm \cite{srivastava2008models} is proven to attain maximum likelihood estimators of the parameters of two dimensional array variate normal distribution. In (\cite{AKDEMIRjalgstat}, \cite{ohlson2011multilinear} and \cite{hoff2011hierarchical}), the flip flop algorithm was extended to general array variate case. 

For the incomplete matrix variate observations with Kronecker delta covariance structure parameter estimation and missing data imputation methods have been developed in \cite{allen2010transposable}.

The following is a modification of the Flip-Flop algorithm for the incomplete array variable observations:

\begin{alg}
\label{alg:1}

Given the current values of the parameters, repeat steps 1 and 2 until convergence:
\begin{enumerate}
    \item Update $\widehat{\widetilde{Y}}_i$ using (\ref{eq:blupforx}),
    \item Update $\widetilde{\mathcal{M}}$ using (\ref{em1}),
    \item For $k=1,2,\ldots,i$ update $\bSigma_k$ using (\ref{em12}).
\end{enumerate}
\end{alg}
In sufficient number of steps, Algorithm \ref{alg:1} will converge to a local optimum of  the likelihood function for the model in \ref{arraymodel}.  In the first step of the algorithm, we calculate the expected values of the complete data  given the last updates of the parameters and the observed data. In the second step, we calculate the value of the mean parameter that maximizes the likelihood function given the expected values of the response and the last updates for the covariance parameters.  In the third step, for each  $k=1,2,...,i,$ the likelihood function for $\bSigma_k$ is concave given the other parameters and the current expectation of the response, i.e., we can find the unique global maximum of this function with respect to $\bSigma_k$ and we take a step that improves the likelihood function. Our algorithm is, therefore, a generalized expectation maximization (GEM) algorithm which will converge to the local optimum of the likelihood function by the results in \cite{dempster1977maximum}.

\section{A semi-parametric mixed effects model}\label{sec5}

A semi-parametric mixed effects model (SPMM)  for the $n\times 1$ response vector $\by$ is expressed as 
\begin{equation}\label{eq:spmm} \by=X\beta+Z\bg+\be \end{equation} where $X\beta$ is the $n\times 1$ mean vector, $Z$ is the $n\times q$ design matrix for the random effects; the random effects $(\bg',\be')'$ are assumed to follow a multivariate normal distribution with mean $\bzero$ and covariance \[ \left( \begin{array}{cc}
\sigma^2_g K  & \bzero  \\
\bzero & \sigma^2_e I_n \end{array} \right)\] where $K$ is  a $q\times q$ kernel matrix. In general, the kernel matrix is a $k\times k$ non-negative definite matrix that measures the known degree of relationships between the $k$ random effects. By the property of the multivariate normal distribution, the response vector $\by$ has a multivariate normal distribution with mean $X\beta$ and covariance $\sigma^2_g(ZKZ'+\lambda I)$ where $\lambda=\sigma^2_e/\sigma^2_g.$

The parameters of this model can be obtained maximizing the likelihood or the restricted likelihood (defined as the likelihood function with the fixed effect parameters integrated out (Dempster 1981) ). The estimators for the coefficients of the SPMM in (\ref{eq:spmm}) can be obtained via Henderson's iterative procedure. Bayesian procedures are discussed in detail in the book by Sorensen \& Gianola. An efficient likelihood based algorithm (the efficient mixed model association (EMMA)) was described in Kang et al. (2007).  

When there are more than one sources of variation acting upon the response vector $\by$, we may want to separate the influence of these sources. For such cases, we recommend using the following multi-way random effects model based on the multi-way normal distribution in Definition \ref{arraymodel}.

\begin{dfn}\label{dfn:AVSPMM}
A multi-way random effects model (AVSPMM) for the $m_1\times m_2, \ldots\times m_i$ response array $\widetilde{Y}$ can be expressed as \small
\begin{equation}\label{eq:mspmm} \widetilde{Y}\sim\phi( \widetilde{Y}; \widetilde{\mathcal{M(\bx)}},\sigma (K_1+\lambda_1I_{m_1})^{1/2}, (K_2+\lambda_2I_{m_2})^{1/2}, \ldots, (K_i+\lambda_iI_{m_i})^{1/2})\end{equation}  where \normalsize $\widetilde{M(\bx)}$ is an $m_1\times m_2, \ldots\times m_i$ dimensional mean function of the observed fixed effects $\bx;$  and $K_1,$ $K_2,$ $\ldots,$ $K_i$ are $m_1\times m_1,$ $m_2\times m_2,$ $\ldots, $ $m_i\times m_i,$ dimensional known kernel matrices measuring the similarity of the $m_1,$ $m_2,$ $\ldots,$ $m_i$ levels of the random effects. If the covariance structure along the $j$th dimension is unknown then the covariance along this dimension is assumed to be an unknown correlation matrix, i.e., we replace the term $(K_j+\lambda_jI_{m_j})$ by a single covariance matrix  $\bSigma_j.$

\end{dfn}
The parameter $\sigma$ is arbitrarily associated with the first variance component and measures the total variance in the variable $\widetilde{Y}$ explained by the similarity matrices $K_1,$ $K_2,$ $\ldots,$ $K_i.$  $\lambda_k$ represents the error to signal variance ratio along the $k$th dimension. For the identifiability of the model additional constraints on the covariance parameters are needed. Here, we adopt the restriction that the first diagonal element of the unknown covariance matrices is equal to one.

It is insightful to write the covariance structure for the vectorized form of the 2-dimensional array model: In this case, \begin{eqnarray}cov(rvec(\widetilde{Y}))&=&\sigma^2(K_2+\lambda_2 I_{m_1})\otimes(K_1+\lambda_1 I_{m_2})\nonumber \\
&=&\sigma^2(K_2\otimes K_1+\lambda_1 K_2\otimes I_{m_1}+\lambda_2 I_{m_2}\otimes K_1+\lambda_1\lambda_2I_{m_1m_2}).\end{eqnarray} If the covariance structure along the second dimension is unknown then the model for the covariance of the response becomes \begin{eqnarray}\label{eq:covstruct2}cov(rvec(\widetilde{Y}))&=&\sigma^2(K_2+\lambda_2 I_{m_1})\otimes\bSigma_2 \nonumber \\
&=&\sigma^2(\bSigma_2\otimes K_1+\lambda_1 \bSigma_2\otimes I_{m_1}).\end{eqnarray}

It should be noted that the SPMM is related to the reproducing kernel Hilbert spaces (RKHS) regression so as the AVSPMM. The similarity of the kernel based SPMM's and reproducing kernel Hilbert spaces (RKHS) regression models has been stressed recently (\cite{gianola2008reproducing}). In fact, this connection was previously recognized by \cite{kimeldorf1970correspondence}, \cite{harville1983discussion}, \cite{robinson1991blup} and \cite{speed1991blup}. RKHS regression models use an implicit or explicit mapping of the input data into a high dimensional feature space defined by a kernel function. This is often referred to as the ''kernel trick'' (\cite{scholkopflearning}).

A kernel function, $k(.,.)$ maps a pair of input points $\bx$ and $\bx'$ into real numbers. It is by definition symmetric ($k(\bx,\bx')=k(\bx',\bx)$) and non-negative. Given the inputs for the $n$ individuals we can compute a kernel matrix $K$ whose entries are $K_{ij}=k(\bx_i,\bx_j).$ The linear kernel function is given by $k(\bx; \by) = \bx'\by.$ The polynomial kernel function is given by $k(\bx; \by) =(\bx'\by+ c)^d$ for $c$ and  $d$ $\in$ $R.$  Finally, the Gaussian kernel function is given by $k(\bx; \by) = \frac{1}{\sqrt{2\pi h}}exp(-(\bx'-\by)'(\bx'-\by)/2h)$ where $h>0.$ Taylor expansions of these kernel functions reveal that each of these kernels correspond to a different feature map. 

RKHS regression extends SPMM's by allowing a wide variety of kernel matrices, not necessarily additive in the input variables, calculated using a variety of kernel functions. The common choices for kernel functions are the linear, polynomial, Gaussian kernel functions, though many other options are available.

We also note that the AVSPMM is different than the standard multivariate mixed model for the matrix variate variables (\cite{henderson1976multiple}), in which, the covariance for the vectorized form of the response vector is expressed as 
\begin{eqnarray}cov(rvec(\widetilde{Y}))&=& (\bSigma_{21}\otimes K_1+\bSigma_{22}\otimes I_{m_1})\end{eqnarray} where $\bSigma_{21}$ and $\bSigma_{22}$ are $m_2$ dimensional  unconstrained covariance matrices and the structure in (\ref{eq:covstruct2}) can be obtained by the restriction $\bSigma_{21}=\bSigma_{22}.$

\subsection{Models for the mean}

A simple model for the mean is given by \begin{equation}\label{eq:allenmean}\widetilde{M}=(\beta_1)^1 \bone_{1\times m_2\times m_3 \times \ldots\times m_i}+(\beta_2)^2 \bone_{m_1\times 1 \times m_3 \times \ldots\times m_i}+\ldots+(\beta_i)^i\bone_{m_1\times m_2\times m_3 \times \ldots\times 1} \end{equation} where the $\beta_k\in \mathbf{R}^{m_k}$ for $k=1,2,\ldots,i$ are the coefficient vectors and the notation $\bone_{m_1\times m_2\times m_3 \times \ldots\times m_i}$ refers to an ${m_1\times m_2\times m_3 \times \ldots\times m_i}$ dimensional array of ones.  Note that this can also be written as \begin{eqnarray}\widetilde{M}&=&(\beta_1)^1(\bone_{m_2})^2\ldots (\bone_{m_k})^k 1 \nonumber \\ &+& (\bone_{m1})^1(\beta_{2})^2(\bone_{m_3})^3\ldots (\bone_{m_k}) ^k 1 \nonumber\\ &+& \ldots+(\bone_{m_1})^1(\bone_{m_2})^2\ldots (\bone_{m_{k-1}})^{k-1} (\beta_{k})^k 1. \nonumber \end{eqnarray} Element-wise, this can be written as \[(\widetilde{M})_{q_1q_2\ldots q_i}=(\beta_1)_{q_1}+(\beta_2)_{q_2}+\ldots+(\beta_i)_{q_i}.\] This generalizes the model for the mean of 2 dimensional arrays recommended in Allen and Tibshirani (2010) to the general $i$ dimensional case. For this model, the fixed effects variables $\bx$ are implicitly the effects of levels of the separable dimensions and some of which might be excluded by fixing the corresponding coefficients vector at zero during the modeling stage. 

If an explanatory variable in the form of an $q$ dimensional vector $\bx$ is observed along with each independent replication of the response variable, we can write a more general mixed model by modeling the mean with \begin{eqnarray}\widetilde{M}(\bx; B_1, \ldots, B_i)&=&(B_1)^1(\bone_{m_2})^2\ldots (\bone_{m_i})^i \widetilde{\bx^1} \nonumber \\ &+& (\bone_{m_1})^1(B_{2})^2(\bone_{m_3})^3\ldots (\bone_{m_i}) ^i \widetilde{\bx^2} \nonumber\\ &+& \ldots+(\bone_{m_1})^1(\bone_{m_2})^2\ldots (\bone_{m_{i-1}})^{i-1} (B_{i})^i \widetilde{\bx^i}. \label{eq:newmean}\end{eqnarray} where $B_k$ is $m_k\times q$ for $k=1,2,\ldots, i$ and $\widetilde{\bx^k}$ stands for the $q \times \ldots \times 1 \times 1\times \ldots \times 1$ dimensional array with $q$ elements of $\bx$ aligned along the $k$th dimension. This model encompasses the model for mean in (\ref{eq:allenmean}). At the modeling stage some of $B_k$ can be excluded from the model by fixing it at $\bzero.$

Let $\widetilde{Y}_1,\widetilde{Y}_2,\ldots,\widetilde{Y}_N$ be a random sample of array observations from the distribution with density $\phi(\widetilde{Y}; \widetilde{M}(\bx; B_1, \ldots, B_i),\mathcal{A}_1,\mathcal{A}_2,\ldots \mathcal{A}_i).$ Assuming that all parameters except $B_k$ are known, the variable \[\widetilde{Z}_\ell=(\widetilde{Y_\ell}-\widetilde{M}(\bx_{\ell}; B_1, \ldots, B_{k-1},B_k=0, B_{k+1},\ldots, B_i)\] has density $\phi(\widetilde{Z}_{\ell};$ $\widetilde{M}(\bx_{\ell}; B_1=0, \ldots, B_{k-1}=0,B_k, B_{k+1}=0,\ldots, B_i=0),$ $A_{1},\ldots, A_{i}).$  Let $Z_{(k)\ell}$ denote the $m_k \times \prod_{j\neq k} m_j$ matrix obtained by matricization of $\widetilde{Z}_\ell$ along the $k$th dimension. $Z_{(k)\ell}$  $=(\bz_{1\ell},$ $\bz_{2\ell},$ $...$ $\bz_{\prod_{j\neq k}m_j \ell})$ has a  matrix-variate normal distribution with mean $B_k \bx_\ell \bone'_{\prod_{j\neq k}m_j}$ and covariances $A_k$ and $A_{-k}$ where $A_{-k}=A_i\otimes A_{i-1} \otimes \ldots \otimes A_{k-1} \otimes A_{k+1}\otimes \ldots \otimes A_1.$ Let $Z^*_{(k)\ell}=Z_{(k)\ell}A_{-k}^{-1}$ and $X^*_{(k)\ell}=\bx_\ell \bone'_{\prod_{j\neq k}m_j}A_{-k}^{-1}.$ Using the results that are already available for the multivariate regression (\cite{anderson1984introduction}), we can obtain the maximum likelihood estimator of $B_k;$ \begin{equation}\label{eq:bhat} \widehat{B_k}=\left[\sum_{\ell=1}^N Z^*_{(k)\ell}{{X^{*}}'_{(k)\ell}}\right]\left[\sum_{\ell=1}^N X^*_{(k)\ell}{{X^{*}}'_{(k)\ell}}\right]^{-1}.\end{equation}

Finally, let an explanatory variable in the form of a $1 \times \ldots \times m_j \times 1\times \ldots \times 1\times q$ dimensional array $\widetilde{X}$ is observed with each independent replication of the response variable, we can write a more general mixed model by modeling the mean with \small \begin{eqnarray} & & \widetilde{M}(\widetilde{X}_\ell; B_1, \ldots, B_i) \nonumber \\ &=&(B_1)^1(\bone_{m_2})^2\ldots (\bone_{m_{j-1}})^{j-1}(I_{m_j})^j (\bone_{m_{j+1}})^{j+1}\ldots (\bone_{m_i})^i \widetilde{X}^{1}_\ell \nonumber \\ &+& (\bone_{m_1})^1(B_{2})^2(\bone_{m_3})^3\ldots(\bone_{m_{j-1}})^{j-1}(I_{m_j})^j (\bone_{m_{j+1}})^{j+1}\ldots (\bone_{m_i}) ^i \widetilde{X}^{2}_\ell\nonumber\\ &+& \ldots \nonumber \\ &+& (\bone_{m_1})^1(\bone_{m_2})^2 \ldots(\bone_{m_{j-1}})^{j-1}(I_{m_j})^j (\bone_{m_{j+1}})^{j+1} \ldots (\bone_{m_{i-1}})^{i-1} (B_{i})^i \widetilde{X}^{i}_\ell. \label{eq:newmean}\end{eqnarray} \normalsize where $B_k$ is $m_k\times q$ for $k=1,2,\ldots, i$ and $\widetilde{X}_\ell^{k}$ stands for the $q \times \ldots \times m_j \times 1\times \ldots \times 1$ dimensional array obtained by  stacking $q \times \ldots \times 1 \times 1\times \ldots \times 1$ arrays $\widetilde{\bx_{\ell_c}^k}$ $c=1,2,\ldots, m_j$ along the $j$th dimension.

Let $\widetilde{Y}_1,\widetilde{Y}_2,\ldots,\widetilde{Y}_N$ be a random sample of array observations from the distribution with density $\phi(\widetilde{Y}; \widetilde{M}(\widetilde{X}; B_1, \ldots, B_i),\mathcal{A}_1,\mathcal{A}_2,\ldots \mathcal{A}_i).$ Assuming that all parameters except $B_k$ are known, the variable \[\widetilde{Z}_\ell=(\widetilde{Y_\ell}-\widetilde{M}(\widetilde{X}_{\ell}; B_1, \ldots, B_{k-1},B_k=0, B_{k+1},\ldots, B_i)\] has density $\phi(\widetilde{Z}_{\ell};$ $\widetilde{M}(\bx_{\ell}; B_1=0, \ldots, B_{k-1}=0,B_k, B_{k+1}=0,\ldots, B_i=0),$ $A_{1},\ldots, A_{i}).$  Let $Z_{(k)\ell}$ denote the $m_k \times \prod_{j\neq k} m_j$ matrix obtained by matricization of $\widetilde{Z}_\ell$ along the $k$th dimension. $Z_{(k)\ell}$  $=(\bz_{1\ell},$ $\bz_{2\ell},$ $...$ $\bz_{\prod_{j\neq k}m_j \ell})$ has a  matrix-variate normal distribution with mean $B_k \widetilde{X}^k_\ell B_{-k}$ where $B_{-k}=\bone'_{m_i} \otimes \bone'_{m_{i-1}} \otimes\ldots \otimes\bone'_{m_{k-1}}\otimes\bone'_{m_{k+1}}\otimes\ldots \otimes\bone'_{m_{j-1}}\otimes I_{m_j}\otimes  \bone'_{m_{j+1}}\otimes \ldots \otimes \bone'_{m_{1}}.$ Row and column covariances of $Z_{(k)\ell}$ are given by $A_k$ and $A_{-k}$ where $A_{-k}=A_i\otimes A_{i-1} \otimes \ldots \otimes A_{k-1} \otimes A_{k+1}\otimes \ldots \otimes A_1.$ Let $Z^*_{(k)\ell}=Z_{(k)\ell}A_{-k}^{-1}$ and $X^*_{(k)\ell}=\widetilde{X}^k_\ell B_{-k}A_{-k}^{-1}.$  The maximum likelihood estimator of $B_k$ is given by \begin{equation}\label{eq:bhat} \widehat{B_k}=\left[\sum_{\ell=1}^N Z^*_{(k)\ell}{{X^{*}}'_{(k)\ell}}\right]\left[\sum_{\ell=1}^N X^*_{(k)\ell}{{X^{*}}'_{(k)\ell}}\right]^{-1}.\end{equation} 

$\widehat{B_k}$ is an unbiased estimator for $B_k$ and the covariance of it is given by \[cov(rvec(\widehat{B}_k))=\left[\sum_{\ell=1}^N X^*_{(k)\ell}{{X^{*}}'_{(k)\ell}}\right]^{-1}\otimes A_k.\] 

A natural generalization of tests of significance of regression coefficients in univariate regression for $B_k$ is 
\begin{eqnarray*}
H_0 &: & L_kB_k=\bzero_l \\
H_1 & :&L_kB_k \neq \bzero_l.
\end{eqnarray*}
Letting \[H_{L_k}=(L_k\widehat{B_k})'(L_k\left[\sum_{\ell=1}^N X^*_{(k)\ell}{{X^{*}}'_{(k)\ell}}\right]^{-1}L_k')^{-1}(L_k\widehat{B_k})\] and \[E_k=\sum_{\ell=1}^N (Z^*_{(k)\ell}-\widehat{B_k}X^*_{(k)\ell})(Z^*_{(k)\ell}-\widehat{B_k}X^*_{(k)\ell})'\]
test statistics based on the eigenvalues of $E_kH_{L_k}^{-1}$ can be obtained. Some possibilities are the  multivariate test statistics like Wilk's Lambda, Pillai's Trace, etc,... For these statistics the distribution under the null hypothesis are available and were described in detail in \cite{anderson1984introduction}.

The generalization of the growth curve model to multiway data is obtained by considering form\begin{equation}\label{eq:growth}M(X_{1},X_{2},\ldots, X_{i};\widetilde{B})=(X_1)^1(X_2)^2\ldots (X_i)^i \widetilde{B}.\end{equation} In (\ref{eq:growth}), $X_{k}$ for $k=1,2,\ldots,i$ are  $m_k\times p_k$ known design matrices and  $\widetilde{B}$ is the unknown parameter array of dimensions $p_1\times p_2 \times \ldots \times p_i.$ For example, if the $k$th dimension $m_1\times m_2 \times \ldots \times m_i $ dimensional response variable $\widetilde{Y}$ is reserved for placing observations taken at points $\{x_{k1}, x_{k2}, \ldots  ,x_{km_i}\},$ $X_k$ might be chosen as the design matrix of the $p_1-1$ degree monomials, i.e.,
\[X_k=\left[ \begin{array}{ccccc}
1 & x_{k1} & x^2_{k1} & . & x^{p_1-1}_{k1} \\
1 & x_{k2} & x^2_{k2} & . & x^{p_1-1}_{k2} \\
. & . & . & . & . \\
1 & x_{km_i} & x^2_{km_i} & . & x^{p_1-1}_{km_i} \\\end{array} \right].\]

Let $\widetilde{Y}_1,\widetilde{Y}_2,\ldots,\widetilde{Y}_N$ be a random sample of array observations from the distribution with density $\phi (M(X_{1},X_{2},\ldots, X_{i};\widetilde{B}),\mathcal{A}_1,\mathcal{A}_2,\ldots \mathcal{A}_i).$ The density of the random sample  $\widetilde{Y}=[\widetilde{Y}_1,\widetilde{Y}_2,\ldots,\widetilde{Y}_N]$ written in the form of a $m_1\times m_2 \times \ldots \times m_i \times N$ array is  $\phi(M(X_{1},X_{2},\ldots, X_{i}, \bone_N;\widetilde{B}),\mathcal{A}_1,\mathcal{A}_2,\ldots \mathcal{A}_i, I_N).$ Assuming that all parameters except $\widetilde{B}$ are known, the variable  $\widetilde{Y^*}=(A_1^{-1})^1(A_2^{-1})^2\ldots (A_i^{-1})^i (I_N)^{i+1}\widetilde{Y}$ has a $\phi(M(A_1^{-1}X_{1},A_2^{-1}X_{2},\ldots, A_i^{-1}X_{i} ,\bone_N;\widetilde{B}), I_{m_1}, I_{m_2},\ldots, I_{m_i}, I_N)$ distribution. The log-likelihood function is of the form 
\begin{eqnarray}
\ell(\widetilde{B})&\propto &- \left(rvec(\widetilde{Y^*})-rvec(M(X^*_{1},X^*_{2},\ldots, X_i^*, \bone_N;\widetilde{B})) \right)' \nonumber \\
& & \left(rvec(\widetilde{Y^*})-rvec(M(X^*_{1},X^*_{2},\ldots, X^*_{i}, \bone_N;\widetilde{B})) \right) \nonumber \\
&=& -\left(rvec(\widetilde{Y^*})-\bone_N \otimes X^*_{i}\otimes X^*_{i-1}\otimes \ldots \otimes X^*_{1}rvec(\widetilde{B})) \right)'  \nonumber \\
&& \left(rvec(\widetilde{Y^*})-\bone_N \otimes X^*_{i}\otimes X^*_{i-1}\otimes \ldots \otimes X^*_{1}rvec(\widetilde{B})) \right) \nonumber \\
&=&-rvec(\widetilde{Y^*})'rvec(\widetilde{Y^*})+2rvec(\widetilde{Y^*})' \bone_N \otimes X^*_{i}\otimes X^*_{i-1}\otimes \ldots \otimes X^*_{1}rvec(\widetilde{B}) \nonumber \\
&-&rvec(\widetilde{B})'(\bone'_N \bone_N \otimes {X^*}'_{i}X^*_{i}\otimes {X^*}'_{i-1}X^*_{i-1}\otimes \ldots \otimes {X^*}'_{1}X^*_{1}) rvec(\widetilde{B})\nonumber
\end{eqnarray}

Taking the derivatives  of $\ell (\widetilde{B})$ with respect to  $rvec(\widetilde{B})$ and setting it to zero, we arrive at the normal equations \[ (\bone_N \otimes X^*_{i}\otimes X^*_{i-1}\otimes \ldots \otimes X^*_{1})'rvec(\widetilde{Y^*})=(N\otimes {X^*}'_{i}X^*_{i}\otimes {X^*}'_{i-1}X^*_{i-1}\otimes \ldots \otimes {X^*}'_{1}X^*_{1}) rvec(\widetilde{B}).\] A solution of the normal equations can be expressed as
\begin{equation}
{\widetilde{B}}=(({X^*}'_{1}X^*_{1})^{-1}{X^*}'_{1})^1,({X^*}'_{2}X^*_{2})^{-1}{X^*}'_{2})^2,\ldots, ({X^*}'_{i}X^*_{i})^{-1}{X^*}'_{i})^i(\bone'_N/N)^{i+1})\widetilde{Y^*}.
\end{equation}
Since the Hessian matrix \[\frac{\partial^2 \ell(\widetilde{B})}{\partial rvec(\widetilde{B})\partial rvec(\widetilde{B})'}=-(N\otimes{X^*}'_{i}X^*_{i}\otimes {X^*}'_{i-1}X^*_{i-1}\otimes \ldots \otimes {X^*}'_{1}X^*_{1})\] is negative definite $\widehat{\widetilde{B}}$ maximizes the log-likelihood function. Note that, $\widehat{\widetilde{B}}$ is a linear function of $\widetilde{Y}$ and also has also normal distribution given by \[\phi(\widetilde{B},({X^*}'_{1}X^*_{1})^{-1},({X^*}'_{2}X^*_{2})^{-1},\ldots ({X^*}'_{i}X^*_{i})^{-1}, 1/N).\]

\subsection{Models for the covariance}

If all parameters except $A_k$ are known, the maximum likelihood estimator of $A_k$ under the unstructured covariance assumption is given by (\ref{em12}).

Now, we turn our attention to estimation of the covariance parameters $\{\sigma^2,\lambda_k\}$ for $k=1,2,\ldots, i.$ Assume that the mean and all variance parameters other than $\{\sigma^2,\lambda_k\}$ are known. By standardizing the centered array variable in all but the $k$th dimension followed by matricization along the same dimension and finally vectorization (denote this $n^*=N\prod_{j=1}^i m_j$ vector by $\bz_{(k)}$), we obtain a multivariate mixed model for which estimates for $\{\sigma^2,\lambda_k\}$ can be obtained efficiently by using a slight modification of EMMA algorithm  (\cite{kang2008efficient}).   The distribution of the $\bz_{(k)}$ is  \[\phi_{N\prod_{j=1}^i m_j}(\bzero, \sigma^2  (I_{N\prod_{j\neq k} m_j}\otimes K_k+\lambda_k I)).\] 

Let $H_k=(I_{N\prod_{j\neq k} m_j}\otimes K_k+\lambda_k I).$ The likelihood function is optimized at  \[\widehat{\sigma^2}=\frac{\bz_{(k)}'H_k^{-1}\bz_{(k)}}{N\prod_{j=1}^i m_j}\] for fixed values of $\lambda_k.$ Using the spectral decomposition of $H_k=Udiag(\epsilon_1+\lambda_k,\epsilon_2+\lambda_k,\ldots, \epsilon_{N\prod_{j=1}^i m_j}+\lambda_k )U'$  and letting $\eta=U'\by,$ the log-likelihood function for $\lambda_k$ at $\widehat{\sigma}^2$ can be written as  \small
\begin{eqnarray}\label{eq:loglikarray}l(\lambda)&=&\frac{1}{2}\left[-n^* log\frac{2\pi \bz_{(k)}'H_k^{-1}\bz_{(k)}}{n^*}-log|H_k|-n^* \right]\nonumber \\
&=&\frac{1}{2}\left[n^* log\frac{n^*}{2\pi}-n^*-n^* log(\sum_{i=1}^{n^*}\frac{\eta_i^2}{\epsilon_i+\lambda_k})-\sum_{i=1}^{n^*} log(\epsilon_i+\lambda_k)\right]\end{eqnarray} \normalsize which can be maximized using univariate optimization. An additional efficiency is obtained by considering the singular value decomposition of a Kronecker product: \[A\otimes B =(U_AD_AV'_A)\otimes (U_BD_BV'_B)=(U_A\otimes U_B )(D_A\otimes D_B)(V_A\otimes V_B)'.\] That is, the the left and right singular vectors and the singular values are obtained as Kronecker products of the corresponding matrices of the components. Therefore, we can calculate the eigenvalue decomposition of $H_k$ efficiently using \begin{equation}\label{eq:kroneig}H_k=(I\otimes U_k)(I\otimes (D_k+\lambda_k I))(I\otimes U_k)'\end{equation} where $U_k(D_k+\lambda I)U'_k$ is the eigenvalue decomposition of $K_k+\lambda_k I$ and $U_kDU'_k$ is the eigenvalue decomposition of $K_k.$

If there are two sources of inputs  along a dimension of an array resulting in two kernel matrices $K_1$ and $K_2$ then a simple model for the covariance parameter along that dimension is given by considering a combination of these matrices and a product term \[w_1 K_1+w_2 K_2+w_3 K_1\odot K_2\] where the '$\odot$' stands for the Hadamard product operator, $w_j\geq 0$ for $j= 1,2,3$ and $\sum_{j=1}^{3}w_j=1.$  It is easy to extend this idea to more than two sources of input and a rich family of models is possible by considering only subsets of these terms. Some of the other models for the covariance along a dimension are spherical. factor analytic, auto regressive, compound symmetric,  and Toeplitz covariance models.

Finally, consider the following covariance model for the vectorized form of a $m_1 \times m_2$ dimensional array $\widetilde{Y}:$ \[cov(rvec(\widetilde{Y}))=\sigma^2(K_2\otimes K_1+\lambda_1 K_2\otimes I_{m_1}+\lambda_2 I_{m_2}\otimes K_1+\lambda_3I_{m_1m_2}).\] Since $\lambda_3=\lambda_1*\lambda_2$ is not imposed, the array model for the array $\widetilde{Y}$ can not be expressed as in Definition \ref{dfn:AVSPMM}. The model parameters can be estimated, for example, using maximum likelihood. However, the estimation is computationally demanding since the efficiencies due to the Kronecker delta covariance structure are not available here. 
\subsection{A Flip-Flop alogorithm for estimating the AVSPMM}
Algorithm \ref{alg:1} can be adopted for the AVSPMM as follows:

\begin{alg}

Given the current values of the parameters, repeat steps 1 and 2 until convergence:
\begin{enumerate}
    \item Update $\widehat{\widetilde{Y}}_\ell$ using (\ref{eq:blupforx}) for $\ell=1,2,\ldots,N;$
    \item Update $\widetilde{M}(\bx; B_1, \ldots, B_i)$ using (\ref{eq:bhat}) using the imputed arrays $\widehat{\widetilde{Y}}_\ell$ for $k=1,2,\ldots ,i;$
    \item For $k=1,2,\ldots,i$ update $\sigma, \lambda_k$ using (\ref{eq:loglikarray}) and (\ref{eq:kroneig}) if $K_k$ is known, otherwise use (\ref{em12}) to update $\bSigma_k.$
\end{enumerate} 

\end{alg}

\section{Principal component analysis for array variate random variables}\label{sec7}

Principal components analysis (PCA) is a useful statistical technique that has found applications in fields such as face recognition and image compression, and is a common technique for finding patterns in data of high dimension. The end product of PCA is a set of new uncorrelated variables ordered in terms of their variances obtained from a linear combination of the original variables. 

\begin{dfn} For the $m_1 \times m_2 \times\ldots \times m_i$ dimensional array variate random variable $\widetilde{Y},$ the principal components are defined as the principal components of the  $d=m_1m_2\ldots m_i$-dimensional random vector $rvec(\widetilde{Y}).$ \end{dfn} 

For an array normal random variable $\widetilde{Y}$ with $E(rvec(\widetilde{Y}))=\bzero$ and covariance $cov(rvec(\widetilde{Y}))=\Lambda$ the principal components can be obtained by  considering the eigenvalue decomposition of the covariance, $\Lambda=UDU'.$ The columns of $U$ are called the principal components. And $cov(U'rvec(\widetilde{Y}))=D$ is diagonal and the $j$th diagonal element of $D$ corresponds to the variance of the random variable $U'_jrvec(\widetilde{Y}).$ 

When a random sample $\widetilde{Y}_1,\widetilde{Y}_2\ldots \widetilde{Y}_N$ is observed and $\Lambda$ is unknown, the principal components are calculated using the sample covariance for  $rvec(\widetilde{Y}_1),$ $rvec(\widetilde{Y}_2)$ $\ldots $ $rvec(\widetilde{Y}_N)$ instead of $\Lambda$ since for $N>\prod_{k=1}^{i}m_i$ the sample covariance matrix is a consistent estimator of the covariance parameter.  However, for high dimensional arrays, usually $N<\prod_{k=1}^{i}m_i,$ and the sample covariance is not a consistent estimator of $\Lambda$ since it has at least one zero eigenvalue whereas the parameter $\Lambda$ is positive definite. 

If we assume that the variable $\widetilde{Y}$ has a Kronecker delta covariance structure, i.e., $\widetilde{Y}\sim\phi($  $\widetilde{\bzero},$ $\mathcal{A}_1,$ $\mathcal{A}_2,$ $\ldots$ $ \mathcal{A}_i),$ then  $\{\lambda(A_r)_{r_j}\}$ are the $m_j$ eigenvalues of $A_rA'_r$ with the corresponding eigen-vectors $\{(\bx_r)_{r_j}\}$ for $r=1,2,\ldots,i$ and $r_j=1,2,\ldots,m_r,$ then $\Lambda=(A_1A'_1\otimes A_2A'_2\otimes^iA_iA'_i)$ will have  eigen-values $\{\lambda(A_1)_{r_1}\lambda(A_2)_{r_2}\ldots\lambda(A_i)_{r_i}\}$ with corresponding eigen-vectors $\{(\bx_i)_{r_i}\otimes(\bx_2)_{r_2}\otimes\ldots \otimes(\bx_i)_{r_i} \}.$ 

If the covariance parameters parameters $\mathcal{A}_1,$ $\mathcal{A}_2,$ $\ldots$ $ \mathcal{A}_i$ are unknown, we can obtain sample based estimates of them when $N\prod_{j\neq k}m_j>m_k$ (assuming there are no missing cells) using Algorithm \ref{alg:1}.  When covariance components along some of the dimensions are assumed known the criterion for the sample size is further relaxed.  We can estimate the eigenvalues and eigen-vectors of the covariance of $rvec(\widetilde{X})$ by replacing  the parameters by their estimators. 

\section{Illustrations}\label{sec7}

Two real and to simulated data sets are used in this section to illustrate our models. These examples also serve to show the effects of changing sample size, missing data proportion and array dimensions on the performance of  methods. 
\begin{ex}
For this first example, we have generated a random sample of $10\times 4\times 2$ dimensional array random variables according to a known array variate distribution. After that, we have randomly deleted a given proportion of the cells of these arrays. The algorithm for estimation 1 was implemented to estimate the parameters and to impute the missing cells. Finally, the correlation between the observed values of the missing cells and the imputed values and the mean squared error (MSE) of the estimates of the overall Kronecker structured covariance matrix is calculated. We have tried sample sizes of $20, 50$ and $100$ and the missing data proportions of $.4, .3,.2$ and $.1.$ The correlations and the MSE's were calculated for 30 independent replications, and these results are presented in Figure  \ref{fig:1}. As expected, the solutions from our methods improve as the sample size increase or when the proportion of missing cells decrease.

\begin{figure}[htbp]
    \centering
        \includegraphics[angle=270, width=0.7\textwidth]{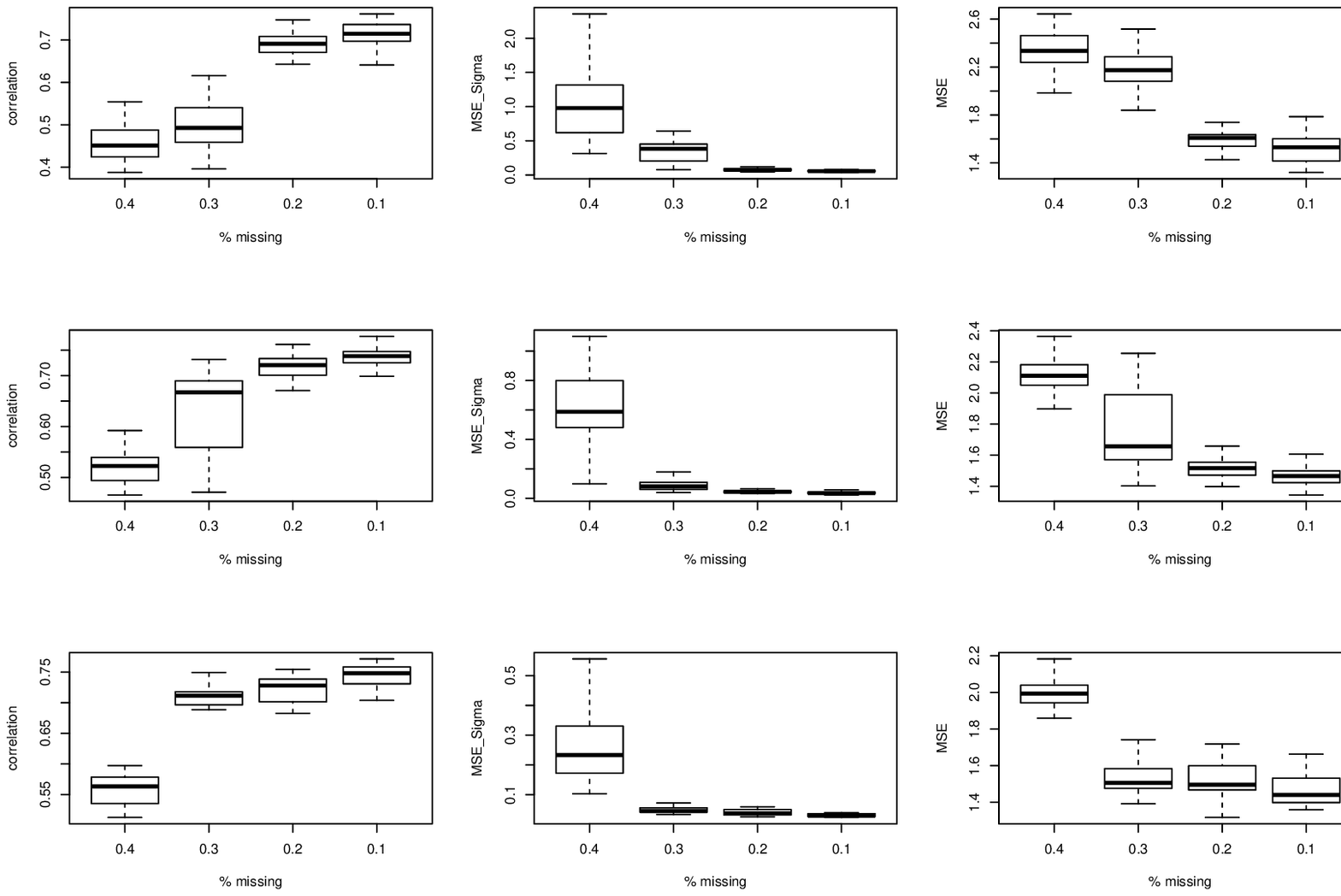}
    \caption{The boxplots of the correlations (left) and the MSEs (right) for varying values of the sample size and missing cell proportions. As expected the solutions from our methods improve as the sample size increase (top to bottom) or when the proportion of missing cells decrease (left to right).}
\label{fig:1}
\end{figure}
\end{ex}

\begin{ex}
In an experiment conducted in Aberdeen during 2013,  524 barley lines from the North American Small Grain Collection were grown using combinations of two experimental factors. The levels of the first factor were the low and normal nitrogen, and the levels of the second experimental factor were dry and irrigated conditions. The low nitrogen and irrigation combination was not reported. Five traits, i.e., plant height, test weight, yield, whole grain protein and heading date (Julian) were used here. We have constructed an incomplete  array of dimensions $524\times 2 \times 2 \times 5$ from this data and induced additional missingness   by randomly selecting a proportion ($.6, .4, .1$) of the cells at random and deleting the recorded values in these cells (regardless of whether the cell was already missing). In addition, 4803 SNP markers were available for all of the 524 lines which allowed us to calculate the covariance structure along this dimension, the covariance structure along the other dimensions were assumed unknown.   An additive mean structure for the means of different traits was used, and all the other mean parameters related to the other dimensions were assumed to be zero.  For each trait, the correlation between the observed and the corresponding estimated values was calculated for 30 independent replications of this experiment  with differing proportion of missing values and these are summarized in Figure \ref{fig:2}. The results indicate that our methods provide a means to estimate the traits that were generated by the combined effect of genetics and environment.

\begin{figure}[htbp]
    \centering
        \includegraphics[angle=270, width=.92\textwidth]{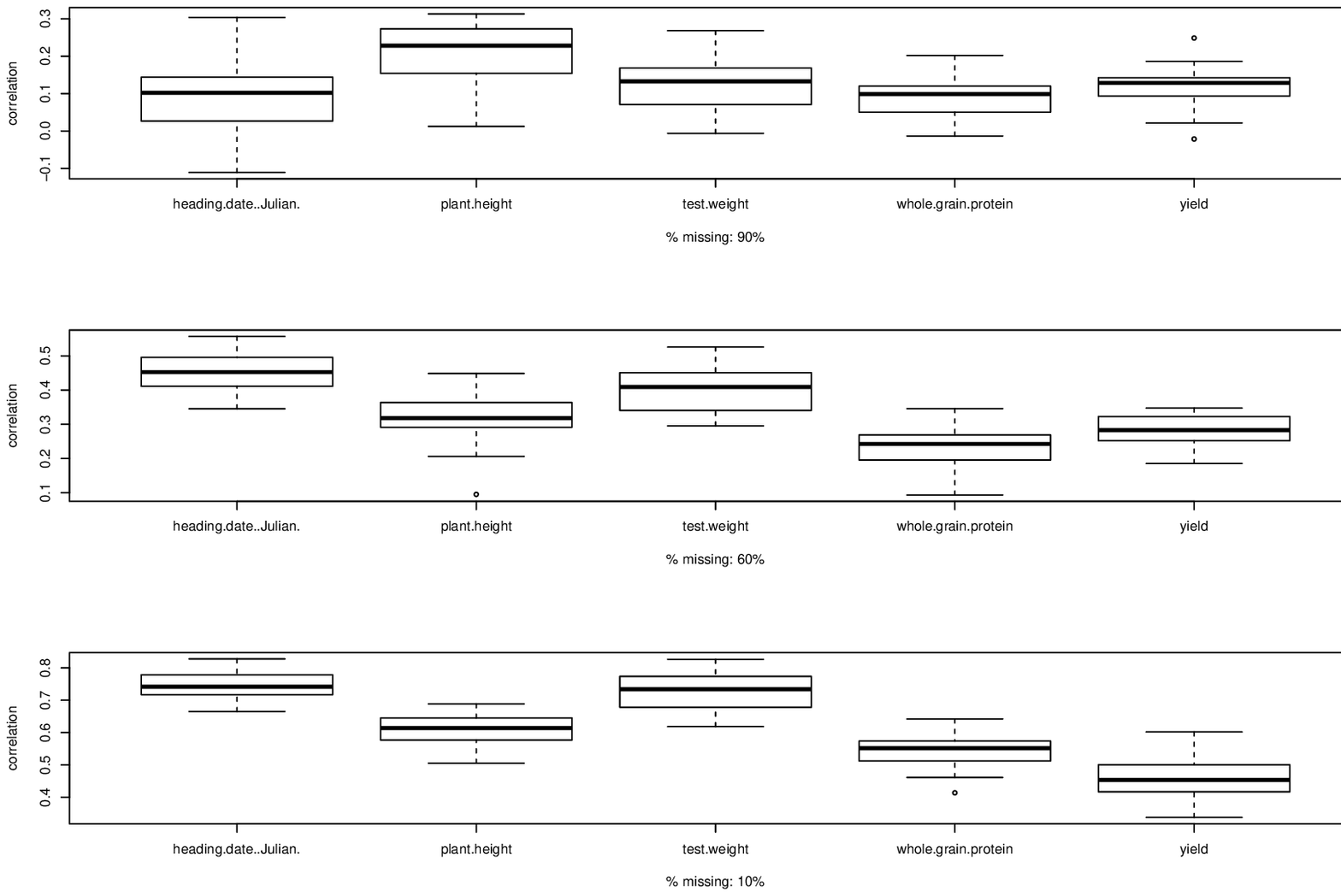}
    \caption{The accuracies for the scenario in Example 2 summarized with the boxplots. The number of missing cells is highest for the bottom figure and lowest for the top figure.}
    \label{fig:2}
\end{figure}

\end{ex}

\begin{ex} 
In this example, we have used the data from an experiment conducted over two years.  365 lines from the spring wheat assocation mapping panel were each observed for three agronomical traits( plant height,    yield, physiological maturity date) in two seperate year/location combinations under the irrigated and dry conditions. A $365 \times 365$ relationship matrix was obtained using 3735 SNP markers in the same fashion as Example 2.  However, since we wanted to study the effect of the number of different genotypes on the accuracies we have selected a random sample of $p_1$ genotypes out of the 365 where $p_1$ was taken as one of $50, 100, 200.$ The phenotypic data was used to form a $p_1\times 2 \times 2 \times 3$ array.  The entry in each cell as deleted with probabilities $.4, .2$ and $.1.$ Finally,  within trait correlations between the missing cells and the corresponding estimates from the AVSPMM over 30 replications of each of the settings of this experiment are summarized by the boxplots in Figure \ref{fig:3}.
\begin{figure}[htbp]
    \centering
        \includegraphics[angle=270, width=.93\textwidth]{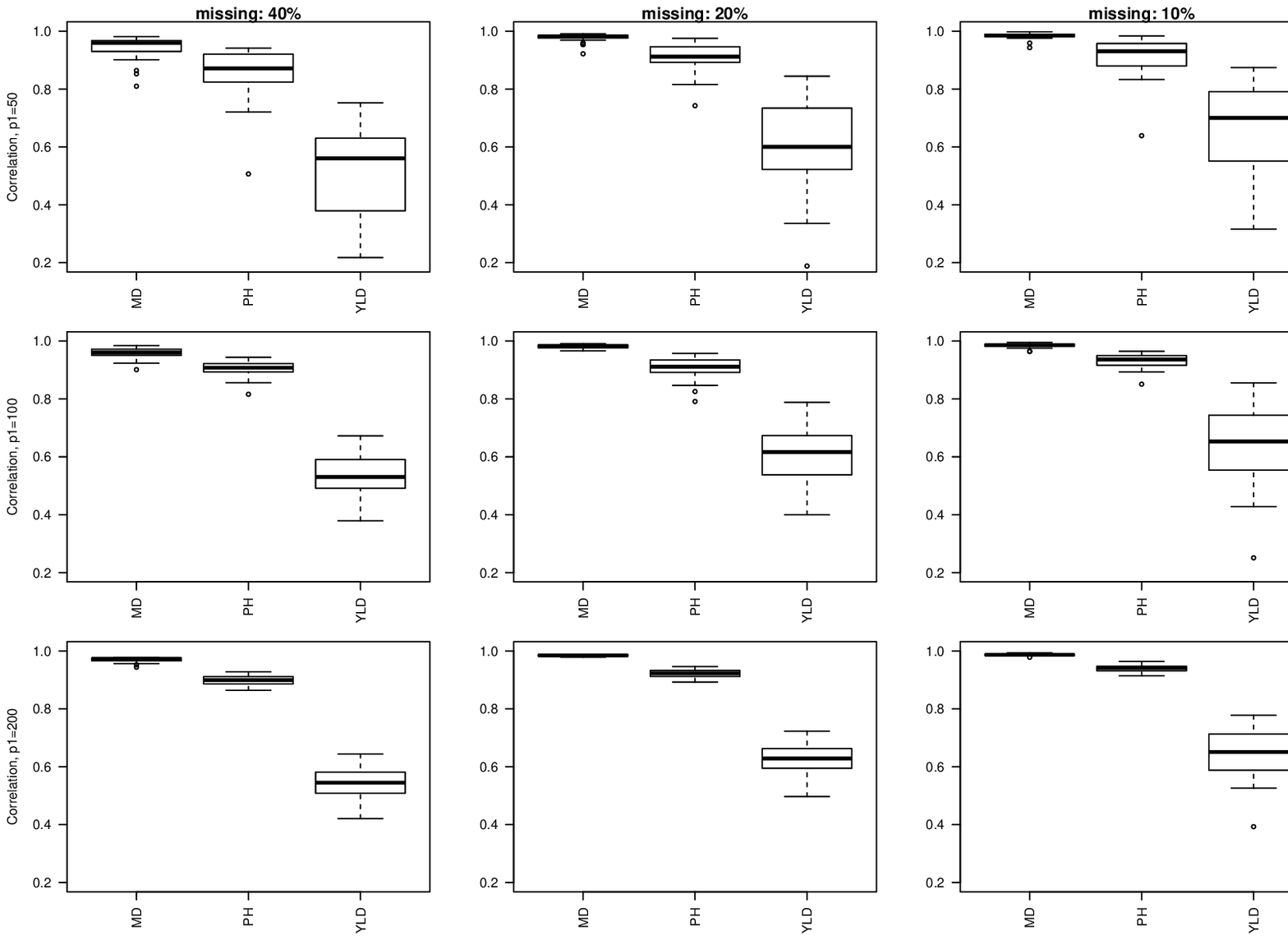}
    \caption{The accuracies for the scenario in Example 3 summarized with the boxplots. The number of missing cells decreases from left to right, and $p_1$ increases from top to bottom. }
\label{fig:3}
\end{figure}
 \end{ex}

\begin{ex}
This data involves simulations from a known AVSPMM model for a $p_1\times 6 \times 2$ array, sample size $1.$ We demonstrate that the MSE for the overall covariance decreases with increasing  $p_1$ where $p_1$ stands for the number of levels of the dimension for which the covariance structure is available in the estimation process. $p_1\times 6 \times 2$ array, sample size $1.$ After generating the array variate response, we have deleted cells with probability $.4, .2,$ or $.1.$ This was replicated 30 times. The correlations and MSE between the estimated response and the corresponding known (but missing) cells and the MSE between the estimated and the known covariance parameters are displayed in Figure \ref{fig:4}.

\begin{figure}[htbp]
    \centering
        \includegraphics[angle=270,width=.9\textwidth]{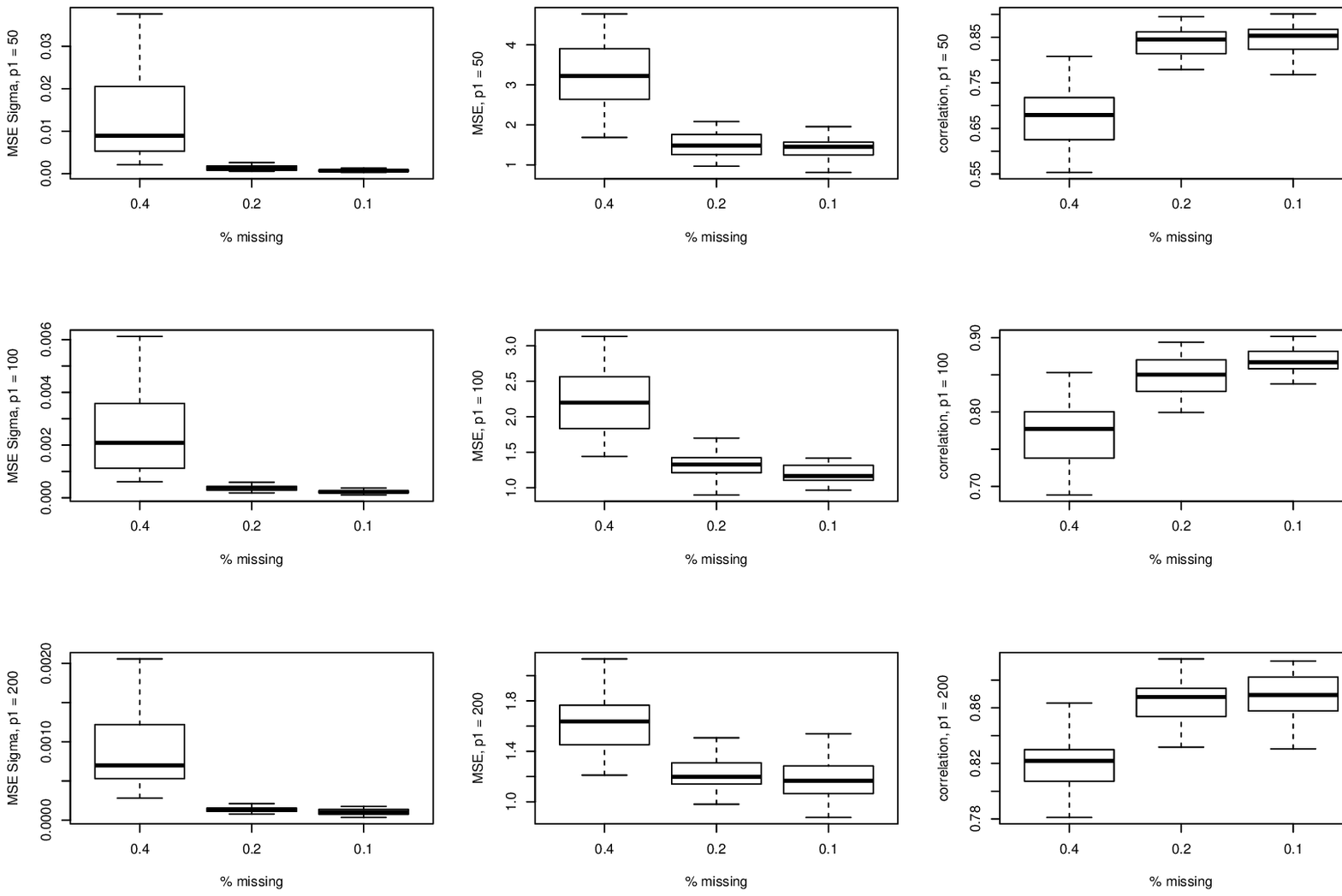}
    \caption{The figures on the left displays the MSE between the estimated and the known covariance parameters and the figures on the right display the correlations between the estimated response and the corresponding known (but missing) cells for $p_1=50, 100, 200$ increasing downwards and probability of missingness $4 , .2, .1.$  decreasing towards the right.}
\label{fig:4}

\end{figure}

\end{ex}

\section{Discussions}\label{sec8}
We have formulated a parametric model for array variate data and developed suitable estimation methods for the parameters of this distribution with possibly incomplete observations. The main application of this paper has been to multi-way regression (missing data imputation), once the model parameters are given we are able to estimate the unobserved components of any array from the observed parts of the array. We have assumed no structure on the missingness pattern; however we have not explored the estimability conditions. 

The proposed algorithms did not always converge to a solution when the percentage of missing values in the array was large. In addition to large percentage of missing values some other possible reasons for non-convergence include poor model specification, the missingness pattern, small sample size, poor initial values for the parameters. In some of the instances of nonconvergence, it might be possible  to obtain convergence by combining the levels of one or more dimensions, and decreasing the order of the array.

Extensions of the AVSPMM are possible by considering other models for the mean and the covariance parameters. Another possible model for the mean array can be obtained by the rank-$R$ decomposition of the mean array parallel factors (PARAFAC) (\cite{harshman1970foundations,bro1997parafac})  where an array is approximated by a sum of $R$ rank one arrays. For a general $i$th order array of dimensions $m_1\times m_2, \ldots\times m_i$ rank-$R$ decomposition can be written as \[\widetilde{M}=\sum_{k=1}^R\rho_k \mu_{r1}\circ \mu_{r2}\circ \ldots\circ \mu_{ri}\] where $\mu_{rk}\in \mathbf{R}^{m_k}$ and $||\mu_{rk}||^2=1$ for $k=1,2,\ldots,i.$ Elementwise, this can be as  \[(\widetilde{M})_{q_1q_2\ldots q_i}= \sum_{k=1}^R\rho_k \mu_{r1q_1}\mu_{r2q_2}\ldots\mu_{riq_i}.\]

The AVSPMM is a suitable model when the response variable is considered transposable. This allows us to separate the variance in the array variate response into components along its dimensions. This model also allows us to make predictions for the unobserved level combinations of the dimensions as long as we know the relationship of these new levels to the partially observed levels along each separate dimension. 

\bibliography{arrayref}
\end{document}